\documentclass[preprintnumbers,nofootinbib,noshowpacs,eqsecnum,prd,superscriptaddress]{revtex4}

\usepackage{graphicx,epsfig}
\usepackage{amsmath,amssymb,bbold,amsbsy,psfrag,color,dsfont,ulem}

\graphicspath{ {Plots/} }

\makeatletter       

\renewcommand{\p@subsection}{}

\makeatother

\def\beq{\begin{equation}}
\def\eeq{\end{equation}}
\def\barr{\begin{array}}
\def\earr{\end{array}}

\begin{document}
\def\lsim{\:\raisebox{-0.5ex}{$\stackrel{\textstyle<}{\sim}$}\:}
\def\gsim{\:\raisebox{-0.5ex}{$\stackrel{\textstyle>}{\sim}$}\:}

\title{ \hfill{\footnotesize{}}                        \\[9mm]
        {\normalsize
        THE EXTENDED HIGGS SYSTEM IN
         \boldmath{$R$}--SYMMETRIC SUPERSYMMETRY THEORIES
} \\[3mm]
}

\author{S.~Y.~Choi}
\address{Department of Physics and RIPC, Chonbuk National University,
               Jeonju 561-756, Korea}
\author{D.~Choudhury}
\address{Department of Physics and Astrophysics, University of Delhi,
               Delhi 110007, India}
\author{A.~Freitas}
\address{Department of Physics \& Astronomy, University of Pittsburgh,
               3941 O'Hara St, PA 15260, USA}
\author{J.~Kalinowski}
\address{Faculty of Physics, University of Warsaw, 00681 Warsaw,
               Poland}
\author{P.~M.~Zerwas}
\address{Institut f\"ur Theoretische Teilchenphysik und Kosmologie,
               RWTH Aachen University, D-52074 Aachen, Germany, and \\
               Deutsches Elektronen-Synchrotron DESY, D-22603 Hamburg, Germany}

\date{\today}

\begin{abstract}
  {\it \noindent The Higgs sector is extended in $R$-symmetric supersymmetry theories
       by two iso-doublets $R_{d,u}$ which complement the standard
       iso-doublets $H_{d,u}$. We have analyzed masses and interactions
       of these novel states and describe their [non-standard] decay
       modes and their production channels at the LHC and $e^+e^-$ colliders.
  }
\end{abstract}

\maketitle

\mbox{ } \\[-1.5cm]

\section{Introduction}
\label{sec:introduction}

\noindent
$R$-symmetry \cite{Salam:1974xa,Fayet:1974pd} is taken in this letter as a continuous
extension of the discrete $R$-parity concept. When this symmetry is imposed on the
Minimal Supersymmetric Standard Model (MSSM) \cite{GomezBock:2007hp,Djouadi:2005gj},
incorporating supersymmetry and electroweak symmetry breaking, some of the basic
parameters are removed from the theory, {\it cf.} Ref.$\,$\cite{Kribs:2007ac}.
Majorana gaugino masses are forbidden, and so are the $\mu$-term and higgsino masses.
No $A$-terms are allowed and so the $L$- and $R$-squark and slepton
mass mixing is absent. In addition the symmetry forbids baryon- and lepton-number
changing terms in the superpotential, as well as dimension-five operators mediating
proton decay \cite{Sakai:1981pk,Weinberg:1981wj}, while Majorana neutrino masses can
be generated. However, by complementing the MSSM gauge supermultiplets
$\hat{G} = \{ G_\mu,\tilde{G} \}$ by additional chiral supermultiplets
$\hat{\Sigma} = \{ {\tilde{G}}',\sigma \}$ in the adjoint representations of the
gauge symmetries, the two gaugino fields can be combined to a Dirac field,
$\tilde{G} \oplus {\tilde{G}}' = {\tilde{G}}_D$, while preserving the $R$-symmetry
for non-zero Dirac masses \cite{Hall:1990hq}.
Complementing the MSSM Higgs sector ${\hat{H}}_{d,u}$ by two iso-doublet fields
${\hat{R}}_{d,u}$ allows one, analogously, to introduce $R$-symmetric $\mu$-terms
and the corresponding higgsino masses.

The transition from Majorana gauginos to Dirac gauginos has far reaching
consequences on supersymmetric particle production at the LHC and $e^+e^-$
colliders, Refs.$\,$\cite{Nojiri:2007jm,Choi:2008pi,Choi:2010gc}, suppressing
pair production channels and cascade decays which are realized in Majorana theories
but forbidden in Dirac theories. $S$-wave $LSP$ neutralino pair annihilation
channels to light fermions are open in collisions of Cold Dark Matter particles
\cite{Hsieh:2007wq, Belanger:2009wf, Chun:2009zx, Chun:2010hz}, while other scenarios
are studied in Ref.$\,$\cite{Kribs:2008hq}. For TeV-scale gauginos, mechanisms for
flavor-changing processes can be suppressed in Dirac theories, alleviating
the flavor problem of supersymmetric theories in models other than gauge
mediation \cite{Kribs:2007ac}. [Additional aspects have recently been discussed in
Refs.$\,$\cite{Benakli:2010gi, Benakli:2009mk}.]

This letter is focussed on the extended $R_{d,u}$-Higgs sector of $R$-symmetric
theories, of which the $R$-Higgs bosons are assigned the conserved $R$-charge $R = 2$.
The mass matrices are derived and the properties of the mass eigenstates are determined.
The dominant interaction terms of the states are identified. Driven by the high $R$-charge,
the properties of the
$R$-Higgs bosons are quite distinct from the ordinary $H$-Higgs bosons, rendering
this system very interesting to explore. From the masses and vertices the decay modes
and production channels can be predicted, building up the phenomenological picture of
the $R$-symmetric Higgs sector which can experimentally be investigated at the LHC
and $e^+e^-$ colliders.

\section{THEORETICAL BASIS}
\label{sec:theoretical_basis}
\noindent
The spectrum of fields in the $R$-symmetric supersymmetry (SUSY)
theory which we will analyze in this letter, consists of the standard
MSSM matter, Higgs and gauge superfields augmented by chiral
superfields $\hat{\Sigma}$ in the adjoint representations
of the corresponding gauge groups and two Higgs iso-doublet superfields
${\hat{R}}_{d,u}$. The $R$-charges of the superfields and their component
fields are listed in Tab.$\,$\ref{tab:Rcharges} [with $R$-charges
$+1/-1$ assigned to the fermionic coordinates $\Theta / \bar{\Theta}$].

\begin{table}[t]
\begin{tabular}{|c|l|l||l|l|l|l|}
\hline
\multicolumn{1}{|c|}{Field} & \multicolumn{2}{c||}{Superfield} &
                              \multicolumn{2}{c|}{Boson} &
                              \multicolumn{2}{c|}{Fermion} \\
\hline \hline
 Matter   &\, $\hat{L}, \hat{E}^c$                    \,& \,\;+1 \,
          &\, $\tilde{L},\tilde{E}^c$                 \,& \, +1 \,
          &\, $L,E^c$                                 \,& $\;\;\,$\,\;0 \,    \\
          &\, $\hat{Q},{\hat{D}}^c,{\hat{U}}^c$       \,& \,\;+1 \,
          &\, $\tilde{Q},{\tilde{D}}^c,{\tilde{U}}^c$ \,& \, +1 \,
          &\, $Q,D^c,U^c$                             \,& $\;\;\,$\,\;0 \,    \\
 $H$-Higgs    &\, ${\hat{H}}_{d,u}$   \,& $\;\;\,$\, 0 \,
          &\, $H_{d,u}$               \,& $\;\;\,$\, 0 \,
          &\, ${\tilde{H}}_{d,u}$     \,& \, $-$1 \, \\
 $R$-Higgs    &\, ${\hat{R}}_{d,u}$   \,& \, +2 \,
          &\, $R_{d,u}$               \,& \, +2 \,
          &\, ${\tilde{R}}_{d,u}$     \,& \, +1 \, \\
 Gauge Vector    &\, $\hat{G}$        \,& \, $\;\,$ 0 \,
          &\, $G_\mu$                 \,& \, $\;\,$ 0 \,
          &\, $\tilde{G}$             \,& \, +1 \,  \\
 Gauge Chiral  &\, $\hat{\Sigma}$     \,& \, $\;\,$ 0 \,
          &\, $\sigma$                \,& \, $\;\,$ 0 \,
          &\, ${\tilde{G}}'$          \,& \, $-$1 \,  \\
\hline
\end{tabular}
\caption{\it The $R$-charges of the superfields and the corresponding bosonic and
             fermionic components.
        }
\label{tab:Rcharges}
\end{table}

With these assignments the gauge, matter/gauge and Higgs/gauge actions
of the MSSM are $R$-invariant, and so is the action associated with the
trilinear Yukawa terms $y_d {\hat{H}}_d \cdot {\hat{Q}} {\hat{D}}^c$ {\it etc}
in the superpotential, the $R$-symmetry being preserved when the electroweak
symmetry is broken. In contrast, the action associated with the standard
$\mu$-term would be $R$-charged and non-invariant.

The presence of the new $R$-Higgs superfields ${\hat{R}}_{d,u}$ with $R=2$
however allows bilinear $\mu$-type mass terms of the form
\begin{eqnarray}
\label{eq:superpot1}
\mathcal{W}_{R} = \mu_d\, \hat{H}_d \cdot \hat{R}_d
                    +\mu_u\, \hat{H}_u \cdot \hat{R}_u\,,
\end{eqnarray}
in the superpotential,
as well as trilinear terms for isospin and hypercharge interactions:
\begin{eqnarray}
\label{eq:superpot2}
\mathcal{W}'_{R} = \lambda^{I,Y}_d\,\hat{H}_d\cdot\hat{\Sigma}_{I,Y}\hat{R}_d
                    +\lambda^{I,Y}_u\,\hat{H}_u\cdot\hat{\Sigma}_{I,Y}\hat{R}_u\,.
\label{eq:trilinear_W}
\end{eqnarray}
Both components carry charges $R = +2$ so that the actions,
$S_{\mathcal{W}}^{(\prime)} = \int \! d^2\Theta \, \mathcal{W}_{R}^{(\prime)} + {\rm h.c.}$,
are $R$-invariant. [The dots in the equations denote the antisymmetric
contraction of the SU(2)$_I$ doublet components.] The mixed $\mu$ terms of
$\mathcal{W}_{R}$ can be generated by the Giudice-Masiero mechanism \cite{Giudice:1988yz},
$\int d^4\Theta\, \hat{X}^\dagger/M \; \hat{H}_i \cdot \hat{R}_i $,
when the chiral spurion in the hidden sector $\hat X$, preserving the $R$-symmetry with
charge $R = 2$, develops the vacuum expectation value $\langle\hat  X\rangle=\Theta^2\, F$.
Similarly, the soft-supersymmetry breaking $B_\mu $ term can be generated
by the $R$-symmetric interaction
\begin{equation}
\int d^4\Theta\, \frac{\langle{\hat{X}}^\dagger \hat{X}\rangle}{M^2}\;
                 \hat{H}_u\cdot \hat{H}_d
\ \ \rightarrow \ \
                 B_\mu {H}_u\cdot {H}_d  \,.
\end{equation}
No bilinear coupling of the $R$-Higgs
fields, which would carry $R$-charges $R = 4$, is present in the
$R$-symmetric theory. Thus, the bilinear Higgs coupling destroys the exchange
symmetry between the $H$ and $R$-Higgs fields.

The soft scalar mass terms of the $R$-Higgs fields  can be generated
in the same manner by the $R$-symmetric term
\begin{eqnarray}
\label{eq:softmasses}
\int \! d^4\Theta\, \frac{\langle {\hat{X}}^\dagger {\hat{X}} \rangle\,}{M^2}
                    {\hat{R}}_d^\dagger {\hat{R}}_d
\ \ \rightarrow \ \
 - m^2_{R_d} ( |R^+_d|^2 + |R^0_d|^2 )\quad {\rm etc} \,,
\end{eqnarray}
suggesting that $\mu_{d,u}$, $B_\mu$ and scalar masses
should be similar in size, {\it i.e.} of the order of the supersymmetry
breaking scale.

While Majorana gaugino masses are forbidden,
Dirac gaugino masses are perfectly allowed in an $R$-symmetric theory.
They can be induced by the interaction of gauge and adjoint chiral
superfields with a hidden sector U(1) gauge superfield $\hat{W}'^\alpha$
with $R=1$ which develops a vacuum $D$-term
$\langle {\hat{W}}'^\alpha \rangle = D\, {\Theta}^\alpha $:
\begin{eqnarray}
\int \! d^2 \Theta \,
 \frac{\langle\hat{W}'^\alpha \rangle}{M} \,\hat{G}_\alpha\hat{\Sigma}
                    + {\rm h.c.}
\ \ \rightarrow \ \
  - M^D\, \tilde{G} \tilde{G}'  +\cdots\,,
\label{eq:Diracaction}
\end{eqnarray}
for isospin, hypercharge and color gauginos. Here, $\hat{G}_\alpha$ are
the gauge superfield-strengths with $R=1$.

It may be noted, for the sake of completeness, that bilinear couplings of the
${\hat{W}}'^\alpha$ and $\hat{X}$ fields with the adjoint chiral fields also
generate contributions $ \int\!d^2\Theta\, {\langle\hat{W}'^\alpha
\hat{W}'_\alpha\rangle}/{M^2} \, {\rm Tr}\, \hat{\Sigma}^2$
and $\int\!d^4\Theta\, {\langle\hat{X}^\dagger \hat{X}\rangle}/{M^2} \,
{\rm Tr}\, {\hat{\Sigma}}^2$ to the soft masses of the adjoint scalars
in addition to the usual soft terms
$\int\!d^4\Theta\, {\langle\hat{X}^\dagger \hat{X}\rangle}/{M^2} \,
{\rm Tr}\, {\hat{\Sigma}}^\dagger \hat{\Sigma}$ [the scale parameters $M$
being generally different in the individual interactions]. In the following
we will not study theoretical points such as SUSY breaking mechanisms, but
explore only the phenomenology of the $R$-symmetric low-energy theory.

\section{MASSES}
\label{sec:masses}

\noindent
The mixing among the states renders the Higgs/scalar sector of the $R$-symmetric
theory very complex. However, since the $\sigma$ sector involves an iso-vector,
the associated mass parameters must be large so as to suppress the related vacuum
expectation value, which is bounded stringently by the small deviation of the
$\rho$ parameter from unity, see Refs.$\,$\cite{Belanger:2009wf,Choi:2010gc}.
For the sake of simplicity we shall
assume that all the mass parameters in the $\sigma$ sector are
large{\footnote{Sgluons in the colored $\sigma$ sector can nevertheless be
produced copiously at the LHC, giving rise to resonance and multi-jet final states,
see Refs.$\,$\cite{Choi:2008ub, Plehn:2008ae, Idilbi:2010rs, Han:2010rf}.}},
{\it viz.} at least of TeV order; systematic expansions for the closely related N=1/N=2
Hybrid Model have been worked out in Ref.$\,$\cite{Choi:2010gc} and the formalism
can easily be adjusted to the present analysis. In the same spirit we shall identify
the $\lambda$ couplings in Eq.$\,$(\ref{eq:trilinear_W}) with the gauge couplings
as prescribed by N=2 extensions, $\lambda^I_d = - \lambda^I_u = - g/\sqrt{2}$
and $\lambda^Y_d=\lambda^Y_u = - g'/\sqrt{2}$, just for definiteness.
Genuine parameters in the $H$ and $R$-Higgs sectors are
consistently assumed of similar size.

\subsection{Higgs Sector}

\noindent
The Higgs potential derives from the actions introduced in the preceding section.
Leaving out the $\sigma$ fields, as argued earlier, the neutral part of the potential
is given by the sum of supersymmetric terms and soft SUSY-breaking terms of the
$H$ and $R$-Higgs fields,
\begin{eqnarray}
{\mathcal{V}}^0_{[H,R]}
   &=&\phantom{+} (m^2_{H_d} + \mu^2_d) |H^0_d|^2 + (m^2_{H_u} + \mu^2_u) |H^0_u|^2
       - (B_\mu H^0_d H^0_u + {\rm h.c.}) \nonumber \\
   && +\, (m^2_{R_d} + \mu^2_d) |R^0_d|^2 + (m^2_{R_u} + \mu^2_u) |R^0_u|^2 \nonumber \\
   && + \left|\lambda^I_d H^0_d R^0_d + \lambda^I_u H^0_u R^0_u\right|^2
      + \left|\lambda^Y_d H^0_d R^0_d - \lambda^Y_u H^0_u R^0_u\right|^2    \nonumber \\
   && + \frac{1}{8}(g^2+g'^2)\left(|H^0_d|^2-|H^0_u|^2-|R^0_d|^2+|R^0_u|^2\right)^2\,.
\end{eqnarray}
Two important consequences follow immediately from the specific form
of the Higgs potential, in particular the absence of the mixed $R^0_d R^0_u$ term
as required by $R$-symmetry:
\begin{itemize}
\item For positive coefficients of the bilinear terms\footnote{Except for small mass
      parameters significantly below the electroweak scale, universal positivity of
      all the coefficients follows from requiring electric charge conservation not
      to be broken spontaneously.}, the $R$-Higgs fields,
      $R_{d,u}$, do not develop non-zero vacuum expectation values, {\it i.e.}
      the potential is minimized for vanishing values of the fields;
\item In the bilinear terms the $R$-Higgs fields with $R=2$ do not mix with the other
      scalar fields with $R\neq 2$ so that, in particular, the $R$-Higgs and $H$-Higgs
      states are mutually independent of each other.
\end{itemize}
\noindent
These conclusions remain valid even if the $\sigma$ fields with $R=0$ are not
discarded.

After subtracting the vacuum expectation values $v_{d,u}/\sqrt{2}$ from the $H$-Higgs
fields $H^0_{d,u}$, the coefficients of the bilinear terms in the neutral and the
corresponding charged part of the potential are identified with the masses of the
physical Higgs fields. The procedure follows the standard way for the $H$-Higgs fields
and will not be repeated here. Given the structure of the Higgs potential,
the $2\times 2$ neutral $R$-Higgs mass matrix reads:
\begin{equation}
{\mathcal{M}}^2_{R^0}
                     =
\left[\begin{matrix}
   m^2_{R_d} + \mu^2_d
  + \frac{1}{2}\left(\lambda^{I2}_d +\lambda^{Y2}_d\right) v^2_d
  - \frac{1}{8} g^2_Z (v^2_d-v^2_u)
&  \frac{1}{2} (\lambda^I_d\lambda^I_u-\lambda^Y_d\lambda^Y_u) v_d v_u \\[2mm]
   \frac{1}{2}(\lambda^I_d\lambda^I_u-\lambda^Y_d\lambda^Y_u) v_d v_u
& m^2_{R_u} + \mu^2_u
  + \frac{1}{2}\left(\lambda^{I2}_u +\lambda^{Y2}_u\right)\, v^2_u
  + \frac{1}{8} g^2_Z (v^2_d-v^2_u)
\end{matrix}\right] \,,
\label{eq:neutral_R_mass_matrix}
\end{equation}
with $g^2_Z = g^2+g'^2$,
in the neutral $(R^0_d, R^0_u)$ basis while the closely-related $2\times 2$
charged $R$-Higgs matrix,
\begin{eqnarray}
{\mathcal{M}}^2_{R^\pm}
                     =
\left[\begin{matrix}
   m^2_{R_d}+\mu^2_d+\lambda^{I2}_d v^2_d- \frac{1}{8} g'^2_Z (v^2_d-v^2_u)
&  0                                                                   \\[2mm]
   0
&  m^2_{R_u}+\mu^2_u+\lambda^{I2}_u v^2_u + \frac{1}{8} g'^2_Z (v^2_d-v^2_u)
\end{matrix}\right] \,,
\label{eq:charged_R_mass_matrix}
\end{eqnarray}
with $g'^2_Z = g^2-g'^2$,
must be diagonal in the charged $(R^\pm_d,R^\pm_u)$ basis because the same-sign
charged $R_{d,u}$-Higgs bosons carry opposite $R$-charges. Thus, the $R$-symmetric
theory incorporates four neutral and four charged $R$-Higgs states in addition to
the standard three neutral and two charged MSSM $H$-Higgs states.

\begin{figure}[t]
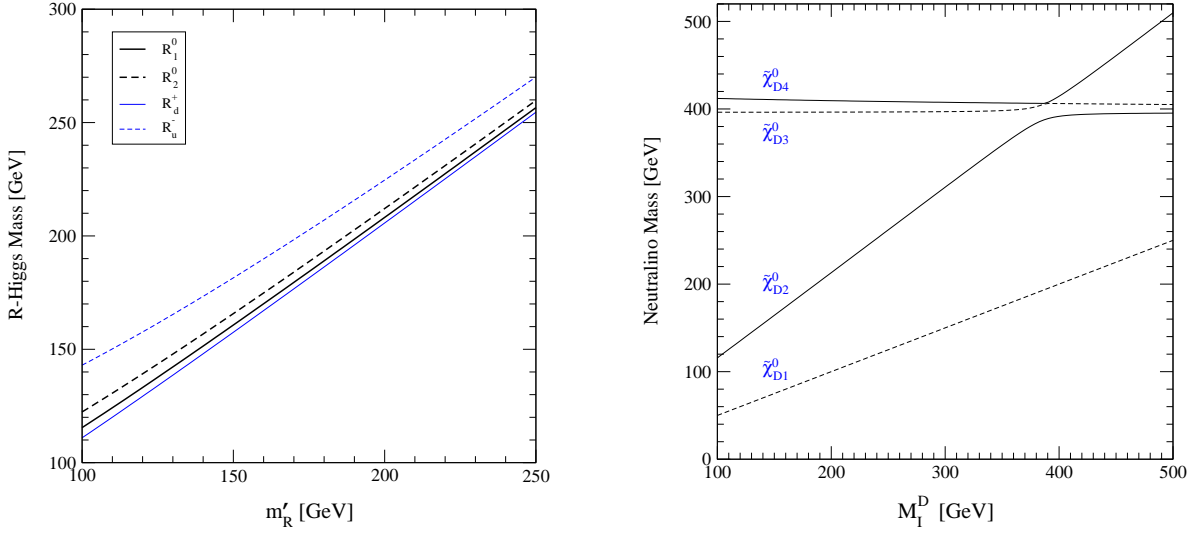

\begin{center}
\epsfig{figure=Rn_Rc_mass.eps, height=7.cm}
\hspace{1cm}
\epsfig{figure=neutralino_mass.eps, height=7.cm}
\end{center}
\caption{\it Left: Masses of the two neutral $R^0_{1,2}$ and the two charged
                   $R^\pm_{d,u}$ Higgs bosons for a common $R$-Higgs mass parameter
                   $m'_R=(m^2_{R_{d,u}}+ \mu^2_{d,u})^{1/2}$ [$M_{R^+_d} \simeq
                   M_{R^0_1}$ as evident from the mass matrices];
             Right: Masses of the four Dirac $\tilde{\chi}^0_{Dj}$ neutralinos
                    for the SU(2)$_I$ Dirac gaugino mass $M^D_I$ [adopting
                    $M^D_Y=M^D_I/2$];
                    the N=2 values of the $\lambda^{Y,I}$ couplings are used
                    and the other parameters are fixed to $\tan\beta=10$ and
                    $\mu_d =\mu_u = 396$ GeV corresponding to the SPS1a$'$
                    point \cite{AguilarSaavedra:2005pw}.}
\label{fig:masses}
\end{figure}

Denoting the diagonal/off-diagonal elements of the neutral $R$-Higgs scalar
mass matrix (\ref{eq:neutral_R_mass_matrix}) by $m'^2_{d,u}$ and $\delta_{du}$,
the corresponding mass-squared eigenvalues read:
\begin{eqnarray}
M^2_{R^0_{1,2}}
   =
\frac{1}{2}\left[m'^2_d + m'^2_u \mp \sqrt{(m'^2_d-m'^2_u)^2
               + 4\delta^2_{du}}\right]\,.
\end{eqnarray}
For mass parameters alike in $d,u$, $(m^2_{R_{d,u}} + \mu_{d,u}^2)^{1/2} \equiv m'_R$,
the neutral and charged masses are degenerate {\it modulo} terms of order $g^2 v^2 / m'_R$,
in analogy to the heavy Higgs bosons of the MSSM.
The $R$-spectrum is displayed
for N=2 couplings $\lambda^{Y,I}$ and mass parameters $m'_R$ in the range between
the approximate LEP limit of about 100 GeV and 250 GeV, where the splitting
is important, in the left frame of Fig.$\,${\ref{fig:masses}}.
The rotation angle between the neutral current and mass eigenstates,
\begin{eqnarray}
\tan 2\alpha_R &=& 2\delta_{du}/(m'^2_d-m'^2_u)\,,
\end{eqnarray}
is generally small, of order $\lambda^2 v^2 / \Delta m'^2$, except for degenerate
$d,u$ mass parameters.

The masses of the $R$-Higgs bosons are restricted only by the non-observation of
their production at LEP and Tevatron, pairwise due to the non-vanishing $R$-charges.
Note however that no dedicated searches for these unusual states, with very
specific decay modes as shown later, have been reported. Moreover, due to the
conserved $R$-charges, the standard electroweak precision observables, $T$ and $S$,
are affected by the novel fields only pairwise at 1-loop order and the contributions
vanish in the limit of degenerate $R$-Higgs masses, in parallel to
Refs.$\,$\cite{Drees:1990dx,Li:1992dt}; the $R$-Higgs masses cannot be
constrained in practice this way.

\subsection{Charginos and Neutralinos}

\noindent
The SUSY partners of the non-colored gauge vector, gauge chiral, $H$-Higgs and
$R$-Higgs bosons generate a rich ensemble of fermionic fields: four charged fields
and their charge-conjugated fields, plus four neutral Dirac fields and their
charge-conjugated fields (see also Ref.$\,$\cite{Kribs:2008hq}).

Fields carrying identical $R$-charges are mixed by the mass matrices, which
emerge from the supersymmetric gauge and Yukawa actions and the supersymmetry-breaking
Dirac actions. They couple the original MSSM gaugino $\tilde{G}$ and higgsino $\tilde{H}$ with
the new gaugino ${\tilde{G}}'$ and higgsino ${\tilde{R}}$ fields, generating
Dirac mass eigenstates $\tilde{\chi}^+_{d1,d2}$, $\tilde{\chi}^-_{u1,u2}$,
$\tilde{\chi}^0_{D1,\ldots\,D4}$ with $R=1$ and their charge-conjugated states with
$R=-1$. Specifically, the two $2\times 2$ chargino mass matrices read
\begin{eqnarray}
{\mathcal{M}}_d^\pm
 &=&
\left[\begin{matrix}
 M^D_I             &  \lambda^I_d v_d   \\
 g v_d/\sqrt{2}    &  -\mu_d
\end{matrix}\right]\,; \quad
{\mathcal{M}}_u^\pm
 =
\left[\begin{matrix}
 M^D_I            & -\lambda^I_u v_u  \\
 g v_u/\sqrt{2}   &  \mu_u
\end{matrix}\right],
\label{eq:chargino_mass_matrices}
\end{eqnarray}
in the $(\tilde{W}'^-,\tilde{H}^-_d)/(\tilde{W}^+,\tilde{R}^+_d)$ basis and
the $(\tilde{W}'^+,\tilde{H}^+_u)/(\tilde{W}^-,\tilde{R}^-_u)$ basis,
respectively, and the $4\times 4$ neutralino mass matrix reads
\begin{eqnarray}
 {\mathcal{M}}^0
 =
\left[\begin{matrix}
 M^D_Y &  0      &  -\lambda^Y_u v_u/\sqrt{2} & \, \phantom{+}\lambda^Y_d v_d/\sqrt{2} \\
   0   &  M^D_I  & -\lambda^I_u v_u/\sqrt{2}  & \, -\lambda^I_d v_d/\sqrt{2}  \\
\phantom{+}g' v_u/2   &  -g  v_u/2  &
 -\mu_u  & 0 \\
-g' v_d/2  & \phantom{+}g  v_d/2    &
   0     & \mu_d
\end{matrix}\right],
\label{eq:neutralino_mass_matrix}
\end{eqnarray}
in the $(\tilde{B}',\tilde{W}'^0,\tilde{H}^0_u,\tilde{H}^0_d)/
(\tilde{B},\tilde{W}^0,\tilde{R}^0_u,\tilde{R}^0_d) $ basis.
The two chargino mass matrices and the neutralino mass matrix,
which is symmetric in the N=2 parametrization of the $\lambda^{I,Y}$ couplings,
can be diagonalized by orthogonal transformations, in parallel to the techniques
used in the MSSM. The physical masses are
generally of the size of the diagonal parameters $M^D_I,M^D_Y,\mu_d,\mu_u$, split up
by corrections of second order in the electroweak scale, $(gv)^2/M_{susy}$.
The mixings are of first order in $g v / M_{susy}$ where $M_{susy}$
denotes the generic SUSY scale. The mass term of the Weyl
bilinears transforms to the Dirac mass term by $M^D (\psi \psi' + {\rm h.c.})
= M^D\, \overline{\psi_D}\,\psi_D$ with $\psi_D =\psi_L + \psi'_R$ and
$\psi'_R = (\psi'_L)^c$. For  Dirac mass terms
varying between 100 GeV and 500 GeV and N=2 values of the $\lambda^{Y,I}$ couplings
the neutralino masses are illustrated in the right frame of Fig.$\,${\ref{fig:masses}}.

\section{DECAY AND PRODUCTION OF \boldmath{$R$}-HIGGS BOSONS}
\label{sec:decay.production}

\noindent
The conserved $R$-charge, $R =2$, of the $R$-Higgs bosons restricts their
trilinear couplings to a small set of exotic sfermion and chargino/neutralino
combinations: the single and double couplings
\begin{eqnarray}
  && R \tilde{\ell} \tilde{\ell} \,,\,
     R \tilde{q} \tilde{q}       \, ;\,
     R \tilde{\chi} \tilde{\chi} \neq 0  \,, \nonumber \\
  && R R H;\, R R V \neq 0\,,
\end{eqnarray}
in symbolic notation. The $R$-Higgs bosons couple only to un-mixed
$L/R$ sfermion pairs and chargino/neutralino pairs transporting $R=2$.
The single $R$-Higgs couplings are generated by $\mathcal{W}_R$ and
$\mathcal{W}'_R$ components of the superpotential, the double $R$-Higgs
couplings by gauge [$V$] and
quartic [$H$] interactions after electroweak symmetry breaking.
Due to the conservation of $R$-charge, the couplings of $R$-Higgs bosons
to pairs of SM-type particles, {\it i.e.} leptons/quarks, gauge bosons
and Higgs bosons, vanish,
\begin{eqnarray}
  Rff;\, RVV;\, RHH = 0 \,,
\end{eqnarray}
[even at loop order, in contrast to the $\sigma$ fields with zero
$R$-charges \cite{Choi:2010gc,Choi:2008ub, Plehn:2008ae, Idilbi:2010rs, Han:2010rf}].
The mass spectrum is taken flavor-diagonal in the sfermion sector but
otherwise most general in the phenomenological analysis.

\subsection{Decay Modes}
\label{subsec:decay}

\noindent
The $R$-Higgs bosons, if kinematically allowed, decay to the following modes, with
the sfermions generated in pairs of $L$ and $R$ particles:
\begin{eqnarray}
R^+_d &\to& \tilde{u}_L \tilde{d}^*_R \ \
            \mbox{and}\ \
            \tilde{\nu}_L \tilde{\ell}^*_R \;\,\,\,
            [\mbox{dimensionful couplings}: \;\, y_d\mu_d\;\,
             \mbox{and}\;\, y_\ell\mu_d] \,, \\[1mm]
R^+_d &\to& \tilde{\chi}^+_{dj} \tilde{\chi}^0_{Dk}
            \;\; [{\rm couplings}: \;\, g,g' \;\, \mbox{for gauge, and}
            \;\, \lambda^{I,Y} \;\, \mbox{for gauge}' \;\, \mbox{components}]     \,.
\end{eqnarray}
The neutral $R_d$ and the charged/neutral $R_u$-Higgs bosons have similar decay channels,
except for the neutral slepton channels. As a result of the vanishing neutrino Yukawa couplings,
they do not decay to up-type sneutrinos. The Yukawa couplings $y$ between $L/R$ sfermions
strongly enhance decays to third-generation channels. The charginos and neutralinos decay,
eventually through cascades, to the lightest stable neutralino $LSP = \tilde{\chi}^0_{D1}$.
To guarantee unstable $R$ Higgs bosons, as assumed in this letter, the $R$-Higgs masses
must exceed twice the $LSP$ mass.

For effective sfermion couplings $\tilde\alpha$ and scalar/pseudoscalar $\tilde{\chi}$ couplings
$\alpha/\alpha'$, the partial on-shell decay widths can be cast into the form:
\begin{eqnarray}
   \Gamma[R \to \tilde{f}_L \tilde{f}'^*_R]
          &=& \frac{\lambda^{1/2}\,{\tilde\alpha}^2_{Rff'} }{16 \pi M_R} \\
   \Gamma[R \to \tilde{\chi}_{Dj} \tilde{\chi}_{Dk}]
          &=& \frac{\lambda^{1/2}}{8 \pi M_R} \,
              \{\alpha^2_{Rjk} [M^2_R - (m_j+m_k)^2] +
                \alpha'^2_{Rjk} [M^2_R - (m_j-m_k)^2] \}  \,,
\end{eqnarray}
with $\lambda = [1-(m_j-m_k)^2/M^2_R][1-(m_j+m_k)^2/M^2_R]$ denoting the
standard phase-space coefficients. The $\alpha$-coefficients are built up
by the mixing matrix elements and the couplings of the current fields.

If none of the visible on-shell channels
is open for kinematic reasons, the $R^0$-Higgs bosons decay visibly to one on-shell
plus one off-shell or to two off-shell states, leading in cascades to final states
$R^0 \to {\tilde{\chi}}^0_{D1} {\tilde{\chi}}^0_{D1} + X^0$, $X^0$ denoting a neutral
set of SM particles. The branching ratios for these visible channels compete with the
exclusive $LSP$ decays, $R^0 \to {\tilde{\chi}}^0_{D1} {\tilde{\chi}}^0_{D1}$,
and will be suppressed in most scenarios.
Charged $R^\pm$-Higgs bosons decay in any case to charged SM particles,
$R^\pm \to {\tilde{\chi}}^0_{D1} {\tilde{\chi}}^0_{D1} + X^\pm$, so that
the charged visible decays are unsuppressed.

For small $R$-Higgs masses the decays $R \to {\tilde{\chi}}^0_{D1} {\tilde{\chi}}^0_{D1} + X$
may also be mediated, though generally at reduced level, by loops interfering with the
tree-level off-shell amplitudes. The loops are built up by chargino/neutralino lines transforming
to the lighter neutralinos by the exchange of $W,Z$ gauge and Higgs bosons,
and radiating the visible markers $X$. Since $L/R$ squarks do not mix in $R$-symmetric
theories, squark/quark triangles emitting pairs of neutralinos/charginos are restricted
effectively to the third generation, with amplitudes of the order of the quark mass
due to chirality flip in the quark line.

\begin{figure}[t]
\begin{center}
\epsfig{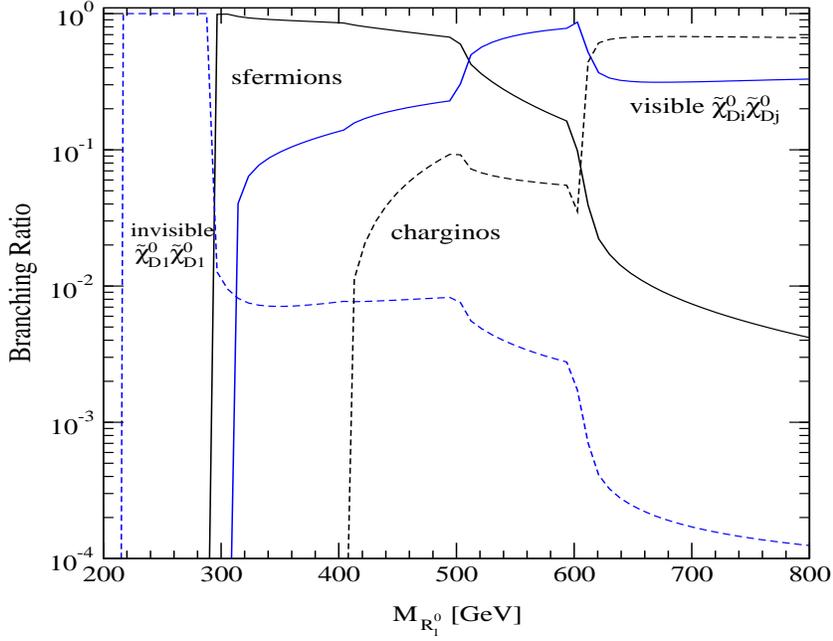}
\end{center}
\caption{\it The leading branching ratios for decays of the neutral
             $R^0_1$-Higgs boson to sfermions, neutralinos and charginos.
             The SPS1a$'$ point \cite{AguilarSaavedra:2005pw} for the sfermion,
             gaugino and higgsino parameters, but without sfermion mixings,
             is adopted for the numerical analysis.}
\label{fig:BR}
\end{figure}

Branching ratios are illustrated for a set of 2-body on-shell neutral
$R^0_1$-Higgs decays in Fig.$\,${\ref{fig:BR}}.

%

\subsection{Production Channels}
\label{subsec:production}

\noindent
Since the $R$-Higgs bosons do not couple to pairs of SM fields,
all standard-type channels are shut for the single production of these
novel particles. Nevertheless, if they are not too heavy, the $R$-Higgs
bosons can be produced in pairs at the
$pp$ collider LHC, via Drell-Yan mechanism, and
at prospective $e^+e^-$ colliders.

\begin{figure}[t]
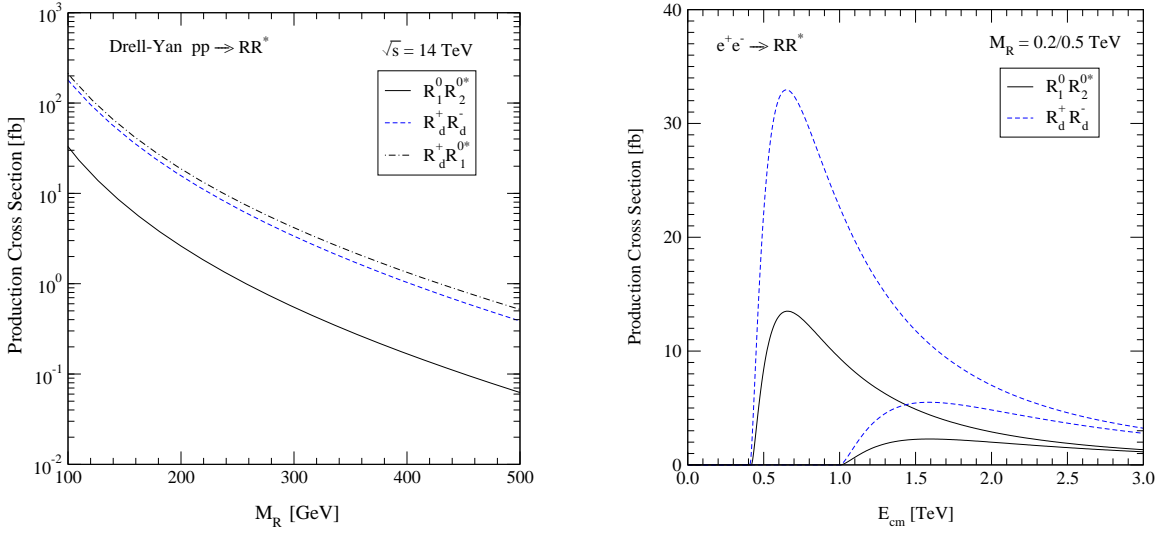

\begin{center}
\epsfig{figure=dyRR.eps, height=7.cm}
\hspace{1cm}
\epsfig{figure=prodRR.eps, height=7.cm}
\end{center}
\caption{\it Left: Drell-Yan production of the neutral/charged $R$-Higgs boson
                   pairs, $R^0_1R^{0{}^*}_2$, $R^+_d R^-_d$ and $R^+_dR^{0{}^*}_1$,
                   at the LHC ($\sqrt{s}=14\,{\rm TeV}$).
                   The cross sections are plotted versus the averaged mass of the
                   produced particles denoted generically by $M_R$;
             Right: Production of the neutral and charged $R$-Higgs boson
                    pairs $R^0_1R^{0{}^*}_2$ and $R^+_d R^-_d$ at TeV
                    $e^+e^-$ colliders [ILC/CLIC]. Two values,
                    0.2 TeV and 0.5 TeV, are taken for the masses. }
\label{fig:prod}
\end{figure}

Pair production via Drell-Yan mechanism and $e^+e^-$ annihilation is described
by the cross sections:
\begin{eqnarray}
\sigma[pp \to RR^{{}^*}]
  &=& \sum_{q\bar{q}}
      \left\langle\,\frac{\pi\lambda^{3/2}}{9s}
          \bigg|\sum_V \alpha_{RRV}\, \frac{s}{s-m^2_V}\, \alpha_{qqV}\bigg|^2\,
      \right\rangle_{q\bar{q}}\,, \\
\sigma[e^+e^- \to RR^{{}^*}]
  &=& \frac{\pi\lambda^{3/2}}{3 s}
      \bigg|\sum_V \alpha_{RRV}\, \frac{s}{s-m^2_V}\, \alpha_{eeV}\bigg|^2\,,
\end{eqnarray}
$V$ denoting the intermediate gauge bosons and $\langle\,\, \rangle_{q\bar{q}}$
the convolution with the $q\bar{q}$ structure functions \cite{Pumplin:2002vw}.
The typical size of the cross sections is illustrated for the
neutral/charged $R$-Higgs pairs, $R^0_1R^{0{}^*}_2$, $R^+_d R^-_d$ and
$R^+_dR^{0{}^*}_1$, in Fig.$\,${\ref{fig:prod}}. [The cross sections for diagonal
neutral $R$-Higgs boson pairs, $R^0_1 R^{0{}^*}_1$ and $R^0_2 R^{0{}^*}_2$, vanish
for the common $R$-Higgs mass parameter $m'_R=(m^2_{R_{d,u}} +\mu^2_{d,u})^{1/2}$
and $N=2$ values of the $\lambda^{I,Y}$ couplings.]
While the Drell-Yan mechanism is useful for moderate $R$-Higgs mass values, {\it cf.}
Ref.$\,$\cite{Alv}, the
$e^+e^-$ collider CLIC extends the mass range into the TeV region.

Additional production mechanisms are provided, with cross sections reduced to
the level of 10 fb for Higgs masses of 100 GeV though,
by fusion channels to Higgs pairs, {\it nota bene} $pp \to \gamma\gamma \to R^+R^-$
\cite{Ohnemus:1993qw,Drees:1994zx,deFavereaudeJeneret:2009db}.

The production of $R$ pairs can also be expected in
heavy MSSM Higgs decays $H \to RR^{{}^*}$, {\it cf.}
$\Gamma [H \to RR^{{}^*}] = {\alpha^2_{RRH}\, \lambda^{1/2}}/{16 \pi M_H}$
with branching ratios up to the level of ten per-cent,
depending in detail on the mixing parameters.

The double production of $R$-Higgs bosons in collider experiments determines the
main characteristics of signatures which can be exploited to search
for these novel states. Though comprehensive analyses
are beyond the scope of this letter, a specific example may nevertheless illuminate
the essential points. We will focus on $R^0$ cascades inspired by the SPS1a$'$ scenario,
in which $\tau'$s are the dominating visible cascade components,
{\it cf.} Ref.$\,$\cite{Martyn}: $R^0 \to {\tilde{\chi}}^0_{D1} {\tilde{\chi}}^0_{D2}$
followed by ${\tilde{\chi}}^0_{D2} \to \tau \tilde\tau$ followed by $\tilde\tau \to
\tau {\tilde{\chi}}^0_{D1}$, {\it i.e.},
\begin{equation}
R^0R^{0{}^*} \to \tau^+ \tau^- \tau^+ \tau^-
                + {\tilde{\chi}}^0_{D1} {\tilde{\chi}}^0_{D1}
                  {\tilde{\chi}}^{0c}_{D1} {\tilde{\chi}}^{0c}_{D1} \,.
\end{equation}
Thus, the final states are characterized by two outstanding signatures:
(i) the high lepton multiplicity of four $\tau'$s, and (ii) four invisible
neutralino $LSP'$s. The $\tau'$s give rise to $e,\mu$ leptons and
pencil-like hadronic jets. The four $LSP'$s generate a large amount of
missing energy in $e^+ e^-$ collisions and missing transverse momentum
in proton collisions. The $\tau$ and missing-energy/momentum distributions
depend characteristically on the masses involved and can discriminate between
signals and backgrounds. In the mass scenario $M_R \sim 2 M_{{\tilde{\chi}}^0_{D2}}
\sim 2 M_{\tilde\tau} \sim 4 M_{{\tilde{\chi}}^0_{D2}}$
the total missing energy amounts approximately to $E_{\perp,miss} \sim 2 M_R$,
while the missing transverse momentum adds up approximately to $p_{\perp,miss}
\sim M_R$, both slightly supplemented by neutrino energies.
Similar final states with multiple $\tau$ leptons and missing energy are
generated, {\it mutatis mutandis}, by other charge configurations and decay
channels of the $R$ pairings. Thus, the multi-fold $\tau$-multiplicity in
association with high values of missing energy/transverse momentum offers
promising signatures for detecting $RR$ events.

\section{SUMMARY}
\label{sec:summ}

\noindent
$R$-symmetry is the basis for an exciting extension of the Minimal Supersymmetric
Standard Model, motivated by the suppression of unwanted processes at the grand
unification scale, and in some scenarios also at the electroweak scale.
In addition to introducing Dirac gauginos and scalar fields in adjoint gauge
representations, the theory predicts two new $R$-Higgs iso-doublets in parallel
to the two standard $H$-Higgs doublets. The $R$-Higgs bosons carry two units of
conserved $R$-charge, leading to novel phenomena in production and decay of these
Higgs bosons. With standard partons in the initial state the $R$-Higgs bosons can
only be produced in pairs, and they decay, finally, to two $LSP'$s, the lightest
neutralinos in general.

\begin{itemize}
\item The leading production mechanism at the LHC is Drell-Yan production of
      $R$-Higgs pairs, mediated by the standard electroweak gauge bosons:
      $pp \to q \bar{q} \to R R^{{}^*}$.
      This production mechanism is useful for low to moderate-size masses.
      For $R^\pm$-Higgs masses below 250 GeV the cross section exceeds 10 fb so that
      3,000 pairs will be produced for an integrated luminosity $\int \mathcal{L}\, dt
      =$ 300 fb$^{-1}$. In $e^+e^-$ collisions particles with masses up to the
      beam energy can be produced at ILC/CLIC, heavy $R$-Higgs bosons
      with masses up to 1.5 TeV at CLIC.

\item Due to the conserved $R$-charge of $R= 2$, the $R$-Higgs bosons can decay
      only into pairs of supersymmetric particles comprising states with $R=2$.
      Decays exclusively to SM particles with no $R$-charge, in parallel
      to MSSM Higgs bosons, are forbidden. Depending on the mass spectra, the $R$-Higgs
      bosons decay to $L/R$ sfermion pairs and pairs of neutralinos and charginos.
      At the end they cascade down to a pair of $LSP'$s plus SM particles $X$:
      $R \to \tilde{\chi}^0_{D1} \tilde{\chi}^0_{D1} + X$. The widths are of electroweak size
      and the lifetimes are short.
\end{itemize}

This well motivated expansion of the Higgs sector can give rise to exciting
phenomena at the LHC and $e^+e^-$ colliders which are very different from the MSSM.
Searching for these novel $R$-Higgs particles and, if discovered, investigating
their properties will therefore be quite interesting. Detailed discussions
of this task are projected for a future study.

\acknowledgments{}

\noindent The work was partially supported by the Korea Research
Foundation Grant funded by the Korean Government (MOERHRD, Basic
Research Promotion Fund) (KRF-2008-521-C00069), by the
Department of Science and Technology, India, under project number
SR/S2/RFHEP-05/2006, as well as by the National Science Foundation
under grant no. PHY-0854782, and by the Polish
Ministry of Science and Higher Education Grant nos.~N~N202~103838
and N~N202~230337.
PMZ thanks the Institut f\"ur Theoretische Teilchenphysik und
Kosmologie for the warm hospitality extended to him at RWTH Aachen
University.\

\vskip 1.5cm

%


\begin{thebibliography}{99}

\bibitem{Salam:1974xa}
  A.~Salam and J.~A.~Strathdee,
  Nucl.\ Phys.\  B {\bf 87} (1975) 85.

\bibitem{Fayet:1974pd}
  P.~Fayet,
  Nucl.\ Phys.\  B {\bf 90} (1975) 104.

\bibitem{GomezBock:2007hp}
  M.~Gomez-Bock, M.~Mondragon, M.~Muhlleitner, M.~Spira and P.~M.~Zerwas,
  arXiv:0712.2419 [hep-ph].

\bibitem{Djouadi:2005gj}
  A.~Djouadi,
  Phys.\ Rept.\  {\bf 459}, 1 (2008)
  [arXiv:hep-ph/0503173].

\bibitem{Kribs:2007ac}
  G.~D.~Kribs, E.~Poppitz and N.~Weiner,
  Phys.\ Rev.\  D {\bf 78} (2008) 055010
  [arXiv:0712.2039 [hep-ph]].

\bibitem{Sakai:1981pk}
  N.~Sakai and T.~Yanagida,
  Nucl.\ Phys.\  B {\bf 197} (1982) 533.

\bibitem{Weinberg:1981wj}
  S.~Weinberg,
  Phys.\ Rev.\  D {\bf 26} (1982) 287.

\bibitem{Hall:1990hq}
  L.~J.~Hall and L.~Randall,
  Nucl.\ Phys.\  B {\bf 352}, 289 (1991).


\bibitem{Nojiri:2007jm}
  M.~M.~Nojiri and M.~Takeuchi,
  Phys.\ Rev.\  D {\bf 76} (2007) 015009
  [arXiv:hep-ph/0701190].

\bibitem{Choi:2008pi}
  S.~Y.~Choi, M.~Drees, A.~Freitas and P.~M.~Zerwas,
  Phys.\ Rev.\  D {\bf 78} (2008) 095007
  [arXiv:0808.2410 [hep-ph]].

\bibitem{Choi:2010gc}
  S.~Y.~Choi, D.~Choudhury, A.~Freitas, J.~Kalinowski, J.~M.~Kim and P.~M.~Zerwas,
  JHEP {\bf 1008} (2010) 025
  [arXiv:1005.0818 [hep-ph]].

\bibitem{Hsieh:2007wq}
  K.~Hsieh,
  Phys.\ Rev.\  D {\bf 77} (2008) 015004
  [arXiv:0708.3970 [hep-ph]].

\bibitem{Belanger:2009wf}
  G.~Belanger, K.~Benakli, M.~Goodsell, C.~Moura and A.~Pukhov,
  JCAP {\bf 0908} (2009) 027
  [arXiv:0905.1043 [hep-ph]].

\bibitem{Chun:2009zx}
  E.~J.~Chun, J.~C.~Park and S.~Scopel,
  JCAP {\bf 1002} (2010) 015
  [arXiv:0911.5273 [hep-ph]].

\bibitem{Chun:2010hz}
  E.~J.~Chun,
  arXiv:1009.0983 [hep-ph].

\bibitem{Kribs:2008hq}
  G.~D.~Kribs, A.~Martin and T.~S.~Roy,
  JHEP {\bf 0901}, 023 (2009)
  [arXiv:0807.4936 [hep-ph]].
\bibitem{Benakli:2010gi}
  K.~Benakli and M.~D.~Goodsell,
  Nucl.\ Phys.\  B {\bf 840} (2010) 1
  [arXiv:1003.4957 [hep-ph]].

\bibitem{Benakli:2009mk}
  K.~Benakli and M.~D.~Goodsell,
  Nucl.\ Phys.\  B {\bf 830} (2010) 315
  [arXiv:0909.0017 [hep-ph]].

\bibitem{Giudice:1988yz}
  G.~F.~Giudice and A.~Masiero,
  Phys.\ Lett.\  B {\bf 206} (1988) 480.

\bibitem{Choi:2008ub}
  S.~Y.~Choi, M.~Drees, J.~Kalinowski, J.~M.~Kim, E.~Popenda and P.~M.~Zerwas,
  Phys.\ Lett.\  B {\bf 672} (2009) 246
  [arXiv:0812.3586 [hep-ph]].

\bibitem{Plehn:2008ae}
  T.~Plehn and T.~M.~P.~Tait,
  J.\ Phys.\ G {\bf 36} (2009) 075001
  [arXiv:0810.3919 [hep-ph]].

\bibitem{Idilbi:2010rs}
  A.~Idilbi, C.~Kim and T.~Mehen,
  Phys.\ Rev.\  D {\bf 82} (2010) 075017
  [arXiv:1007.0865 [hep-ph]].

\bibitem{Han:2010rf}
  T.~Han, I.~Lewis and Z.~Liu,
  arXiv:1010.4309 [hep-ph].

\bibitem{AguilarSaavedra:2005pw}
  J.~A.~Aguilar-Saavedra {\it et al.},
  Eur.\ Phys.\ J.\  C {\bf 46} (2006) 43
  [arXiv:hep-ph/0511344].

\bibitem{Drees:1990dx}
  M.~Drees and K.~Hagiwara,
  Phys.\ Rev.\  D {\bf 42} (1990) 1709.

\bibitem{Li:1992dt}
  L.~F.~Li,
  Z.\ Phys.\  C {\bf 58}, 519 (1993).

\bibitem{Pumplin:2002vw}
  J.~Pumplin, D.~R.~Stump, J.~Huston, H.~L.~Lai, P.~M.~Nadolsky and W.~K.~Tung,
  JHEP {\bf 0207} (2002) 012
  [arXiv:hep-ph/0201195].

\bibitem{Alv}
  A.~Alves and T.~Plehn,
  Phys.\ Rev.\  D {\bf 71} (2005) 115014
  [arXiv:hep-ph/0503135].

\bibitem{Ohnemus:1993qw}
  J.~Ohnemus, T.~F.~Walsh and P.~M.~Zerwas,
  Phys.\ Lett.\  B {\bf 328} (1994) 369
  [arXiv:hep-ph/9402302].

\bibitem{Drees:1994zx}
  M.~Drees, R.~M.~Godbole, M.~Nowakowski and S.~D.~Rindani,
  Phys.\ Rev.\  D {\bf 50} (1994) 2335
  [arXiv:hep-ph/9403368].

\bibitem{deFavereaudeJeneret:2009db}
  J.~de Favereau de Jeneret {\it et al.},
  arXiv:0908.2020 [hep-ph].

\bibitem{Martyn}
  N.~Ghodbane and H.~U.~Martyn,
  in {\it Proc. of the APS/DPF/DPB Summer Study on the Future of Particle Physics (Snowmass 2001)},
  arXiv:hep-ph/0201233.


\end{thebibliography}
\end{document}